\documentclass[aps,twocolumn,prl,superscriptaddress,amsmath,showpacs,tightenlines]{revtex4}
\usepackage{graphicx}
\usepackage{epsfig}
\usepackage{amssymb}
\usepackage{color}

\begin{document}
%
%+++++++++++++++++++++++++++++++++++++++++++++++++++
%  (a) Cluster state or graph state?
%
\title{Dynamical topological quantum computation using spin pulse control in the Heisenberg model}
\author{Tetsufumi Tanamoto}
\affiliation{Corporate R \& D center, Toshiba Corporation,
Saiwai-ku, Kawasaki 212-8582, Japan}

\author{Keiji Ono}
\affiliation{Low temperature physics laboratory, RIKEN, Wako-shi, Saitama 351-0198,
Japan}

\author{Yu-xi Liu}
\affiliation{Institute of Microelectronics, Tsinghua University, Beijing 100084, China}
\affiliation{Center for Emergent Matter Science, RIKEN, Saitama 351-0198, Japan}
\affiliation{Tsinghua National Laboratory for Information Science and Technology (TNList), Beijing 100084, China}

\author{Franco Nori}
\affiliation{Center for Emergent Matter Science, RIKEN, Saitama 351-0198, Japan}
\affiliation{Department of Physics, The University of Michigan, Ann Arbor, Michigan 48109-1040, USA}

\date{\today}
\begin{abstract}
Hamiltonian engineering is an important approach for quantum information 
processing, when appropriate materials 
do not exist in nature or are unstable.
So far there is no stable material for the Kitaev spin Hamiltonian with anisotropic interactions on a honeycomb lattice
(A. Kitaev, Annals of Physics {\bf 321} 2 (2006)), 
which plays a crucial role in the realization of both Abelian and non-Abelian anyons. 
Here, we show 
how to dynamically realize the Kitaev spin Hamiltonian from the conventional Heisenberg spin 
Hamiltonian using a pulse-control technique. 
By repeating the same pulse sequence, the quantum state 
is dynamically preserved. 
The effects of the spin-orbit
interaction and the hyperfine interaction are also investigated. 
\end{abstract}
\pacs{03.67.Lx, 03.67.Mn, 73.21.La}
\maketitle

\begin{figure}
\begin{center}
\includegraphics[width=6cm]{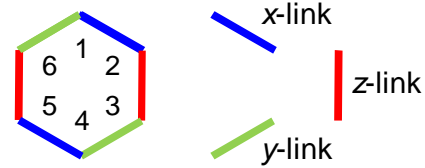}
\end{center}
\caption{(color online) Kitaev model on a honeycomb lattice.} 
\label{honeycomb}
\end{figure}

%\section{Introduction}
Topological quantum computation (TQC) has attracted considerable interest 
due to its robustness to local perturbations~\cite{Wen}.
Anyons, which obey different statistics from bosons and fermions, are also 
of fundamental interest in physics~\cite{Wilczek}.
Kitaev~\cite{Kitaev} provided an important exactly-solvable model of a spin-1/2 system 
on a honeycomb lattice with potential links to topological 
quantum computation, for both Abelian and non-Abelian anyons.
In the Kitaev honeycomb model, spin-spin interactions are realized 
by the so-called $XX$, $YY$, $ZZ$ couplings along three directions (Fig.~\ref{honeycomb}). 
%The anyon properties depend on the relative coupling strength 
%of these interactions.
This spin model stimulated the physics of an 
anyon system, including Majorana fermions.
However, it is not easy to find materials which have such anisotropic interactions.

Even if we can find a possible material for realizing a desired Hamiltonian, 
we have to integrate and fabricate it by attaching many electrodes
and probes to confirm whether it is sufficiently controllable.
Regarding artificial realizations of the Kitaev Hamiltonian,
theoretical proposals have been made using optical lattices~\cite{Duan,Aguado} and superconducting qubits~\cite{You}.
In Ref.\cite{You}, You {\it et al}. used different qubit-qubit interactions 
depending on the coupling direction. 
%but this proposal requires the fabrication over a large area.

The Heisenberg model describes two-body interactions 
in many magnetic materials 
and artificial systems which are constructed  
by various fabrication techniques. 
In particular, quantum dot (QD) systems~\cite{Loss,Burkard,Petta,Burkard2,Xuedong,Ono2},
 donor systems~\cite{Kane,Simmons,Tabe,Ono}, and nitrogen-vacancy (NV) centers~\cite{NV,Jiang,Ping} 
are promising candidates for 
spin-qubit systems, because they could be integrated 
on substrates by using advancing nanofabrication technology. 
The spin qubits in QDs also have the advantage that the tunneling coupling can be varied 
by attaching gate electrodes. 
%For the NV center, the methods of 
%changing the coupling is proposed by the spin-chain model~\cite{}.
Thus, placing QDs, with a single electron each, on a honeycomb lattice site is 
a promising way of realizing a Kitaev spin Hamiltonian.

Here we show how to dynamically generate this Hamiltonian 
starting from the natural Heisenberg model 
by using pulse-control techniques~\cite{Ernst,tanaSW}.
Permanent data cannot be stored in quantum systems
and should be transferred to attached conventional semiconductor memory, 
because the coherence time is less than a microsecond.
%The important data should be quickly exchanged between 
%the quantum information processors and attached conventional semiconductor 
%memory. 
From this point, 
it is natural to treat the qubit states dynamically 
with pulse control.
Here, the Baker-Campbell-Hausdorf (BCH) formula 
is applied for 
producing a Kitaev Hamiltonian by taking into account the direction of the qubit-qubit couplings. 
Because unwanted terms are also generated in the BCH formula,
the engineered Hamiltonian is effective for a finite time interval. 
Therefore, the dynamical
approach requires a {\it refresh} process in which the same 
pulse sequence for generating the Kitaev Hamiltonian 
is carried out. 
The idea of repeating the production process is very common 
in conventional digital computers, such as 
 dynamic random access memory (DRAM),
which 
% which is  a $main$ memory of personal computer. 
%Because the DRAM 
is a big capacitor and the amount of electric charge 
is lost over time~\cite{DRAM}. 
%Therefore a {\it refresh} process is required every 
%less than hundred milliseconds.

%%%%%%%%%%%%%%%%%%%%%%%%%%%%%%%%%%%%%%%%%%%%
%%%%%%%%%%%%%%%%%%%%%%%%%%%%%%%%%%%%%%%%%%%%

Natural qubit-qubit interactions can be described by either Ising, $XY$, or Heisenberg models.
The most feasible one is the Ising interaction and the hardest one is 
the Heisenberg interaction~\cite{tana09}.
It is not easy to transform the Heisenberg interaction to any desirable Hamiltonian 
because Heisenberg interactions have three components.
Thus, the standard way to design Hamiltonians from the Heisenberg interaction
is to change the Heisenberg interaction into an Ising interaction at the first step.
Once we obtain the Ising interaction, we can transform it to a desirable 
interaction. However, as we will show, this method requires more than six steps to 
create a Kitaev Hamiltonian:
three steps for obtaining $XX$,$YY$,$ZZ$ interactions, plus
additional three steps, for obtaining the corresponding $x$, $y$ and $z$-links.
What we would like to show here is that if we carefully design 
two-dimensional (2D) pulses which vary depending on qubit location, 
we can obtain the desired Hamiltonian in only one step.
Because the topological Hamiltonian 
is derived from perturbation theory~\cite{Kitaev},
we have to examine the direct effect of the 
unwanted terms, other than the spin-orbit (SO) terms and the hyperfine(HF) terms, 
on the topological Hamiltonian, focusing on the gapped phase (phase $A$).

%%%%%%%%%%%%%%%%%%%%%%%%%%%%%%%%%%%%%%%%%%%%%%%%%%%%%%%%%%%%%%
%%%%%%%%%%%%%%%%%%%%%%%%%%%%%%%%%%%%%%%%%%%%%%%%%%%%%%%%%%%%%%
%\section{Formulation}
{\it Formulation.---}
The Kitaev Hamiltonian is given by the anisotropic spin model on the honeycomb lattice
\begin{equation}
H_K=-J_x \sum_{x-{\rm links}} X_jX_k -J_y \sum_{y-{\rm links}}Y_jY_k
-J_z \sum_{z-{\rm links}}Z_jZ_k,
\label{Kitaev_H}
\end{equation}
where $X_j$, $Y_j$ and $Z_j$ are the Pauli spin matrices and 
the interaction type  ($x$, $y$, and $z$ links) depends on the direction 
of the bond between the two sites (Fig.~\ref{honeycomb}).
The model in Eq.(\ref{Kitaev_H}) can be mapped to free Majorana fermions
coupled to a $\mathbb{Z}_2$ gauge field and 
have two types of interesting ground states (phase $A$ and phase $B$)
depending on the relative magnitude of $J_x$, $J_y$ and $J_z$.
The region $J_{\alpha_1} \le J_{\alpha_2}+J_{\alpha_3}$, where 
$\alpha_i$ ($i$=1,2,3) refers to $x$,$y$,$z$, is the 
gapless $B$ phase in which non Abelian anyons appear,
and the other region is the gapped phase $A$, where Abelian anyon statistics is expected.  
In the $B$ phase , an additional external magnetic field opens an energy gap.

We would like to derive Eq.~(\ref{Kitaev_H}) from the Heisenberg Hamiltonian given by
\begin{equation}
H_{S}=\sum_{i<j} [J_x X_iX_j+J_y Y_iY_j+J_z Z_iZ_j].
\label{Heisenberg}
\end{equation}
The ``creation'' of Eq.~(\ref{Kitaev_H}) is carried out by combining $H_S$ with 
a transferred Hamiltonian $H_R$, which is produced by applying a customized  
pulse sequence to $H_S$, like NMR, 
using a repetition 
of the BCH formula. 
Concretely, the target Hamiltonian $H_{\rm tgt}$ is obtained by $H_{\rm tgt}=H_S+H_R$, 
such as 
%\begin{equation}
%(e^Ae^B)^n \approx \exp( i t_0 [H_S+H_R] + (t_0^2/ [4 n])[H_S,H_R])
%$(e^{-itH_S}e^{-itH_R})^n \approx \exp( i nt [H_S+H_R] - nt^2[H_S,H_R]+..)$.
$e^{-itH_S}e^{-itH_R} \approx \exp( i t [H_S+H_R] - t^2[H_S,H_R]/2+..)$, 
when $J_\alpha t \lesssim  1$.  %%% \gtrsim
The terms higher than $t$ are the unwanted ones $H_{\rm uw}^{\rm bch}$.
As mentioned in the introduction, the standard way 
is to first create an Ising Hamiltonian and afterwards change the Ising Hamiltonian 
to a Kitaev Hamiltonian. We show that, if we consider the geometric 
distribution of the qubit on the honeycomb lattice, 
we can effectively create the Kitaev Hamiltonian. 

\begin{figure}
\begin{center}
\includegraphics[width=8.5cm]{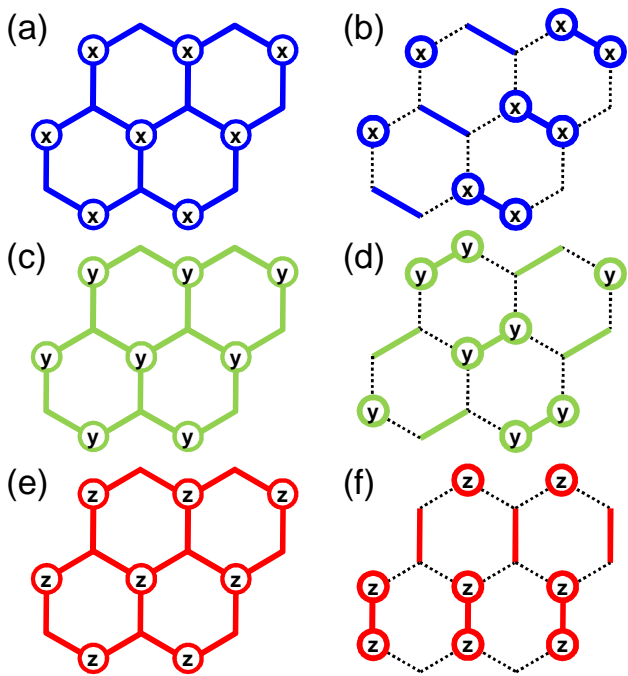}
\end{center}
\caption{(color online) Standard method to dynamically produce a Kitaev Hamiltonian from the Heisenberg model. The symbols $x$, $y$ and $z$ in the lattice sites 
show the application of $\pi/2$-pulses around $x$, $y$ and $z$, respectively.
The bonds with dotted lines indicate that there is no interaction between the connected sites.
(a) Pulse mapping of $P_{1}^x$ to create the Ising Hamiltonian, $H_{\rm step1}^{x}=\sum_{i,j} J_x X_iX_j$ 
in $e^{i t H_{\rm step1}^{x}}=
e^{itH_S}e^{it P_{1}^{x\dagger}H_S P_{1}^x}$.
(b) Pulse mapping to select only the $x$-link of the Kitaev Hamiltonian 
from the Ising Hamiltonian of (a). %, $H_x=\sum_{i,j} J_x X_iX_j$. 
(c) and (e) express pulse distributions for generating 
 $H_y=\sum_{i,j} J_y Y_iY_j$ and  $H_z=\sum_{i,j} J_z Z_iZ_j$, respectively.
(d) and (f) show pulse the pattern to select only the $y$ and $z$ links, respectively.
} 
\label{standard}
\end{figure}

{\it Standard method.---}
The standard way to convert Eq.~(\ref{Heisenberg}) to Eq.~(\ref{Kitaev_H}) 
requires six steps as shown in Fig.~\ref{standard}. 
The first step is to create the three Ising Hamiltonians,
$H_x=\sum_{i,j} J_x X_iX_j$, $H_y=\sum_{i,j} J_y Y_iY_j$, $H_z=\sum_{i,j} J_z Z_iZ_j$, 
from Eq.~(\ref{Heisenberg}) as shown in Figs.~\ref{standard}(a,c,e).
The generated Ising Hamiltonians are described by
$
H_{\rm step1}^{\rm \alpha}=
\log [e^{itH_S}e^{itH_{r1}^\alpha}]/(it),
$
where $H_{r1}^\alpha =P_1^{\alpha \dagger} H_S P_1^{\alpha}$ ($\alpha=x,y,z$) 
is a rotated Hamiltonian by 
applying a $\pi/2$-pulse around the $\alpha$-axes on the lattice sites 
of Figs.~\ref{standard}(a,c,e). 
%%% 20140607 new
The rotations consist of a multiple of the single rotations 
using formula $
e^{-i(\pi/2) \sigma_\alpha}
\sigma_\alpha
e^{i(\pi/2) \sigma_\alpha}
=- \sigma_\alpha$, for $X=\sigma_x$,$Y=\sigma_y$ and $Z=\sigma_z$.
The next step is to erase unnecessary Ising interactions, such as 
$
H_{\rm step2}^{\alpha}= 
\log [e^{itH_{\rm step1}^\alpha }e^{itH_{r2}^\alpha}]/(it),
$
where 
$H_{r2}^\alpha =P_2^{\alpha \dagger}H_{\rm step1}^\alpha P_{2}^\alpha$
is obtained by applying $\pi/2$-pulses 
depending on the links in Figs.~\ref{standard}(b,d,f).
%The Ising interactions for the $y$-links and $z$-links are obtained similarly.
Thus, the Kitaev Hamiltonian is dynamically obtained
by $H_{K}^{\rm std}(t)=\log [e^{it  H_{\rm step2}^{x}}
e^{it  H_{\rm step2}^{y}}e^{it  H_{\rm step2}^z}]/(it)$. 
Note that parts of the Kitaev Hamiltonian do not commute, 
$i.e.$ $[\sum_{x-{\rm links}} X_jX_k, \sum_{y-{\rm links}} Y_jY_k]\neq 0$. Therefore,
the unwanted terms increase in the standard method.
A better method is presented below.  
%In this method, we use the BCH formula six times. As the number of 
%times the BCH formula is increased, the number of unwanted 
%terms $H_{\rm uw}^{\rm bch}$ is increased.
%%%%%%%%%%%%%%%%%%%%%%%%%%%%%%%%%%%%%%%%%%%%

\begin{figure}
\begin{center}
\includegraphics[width=8.5cm]{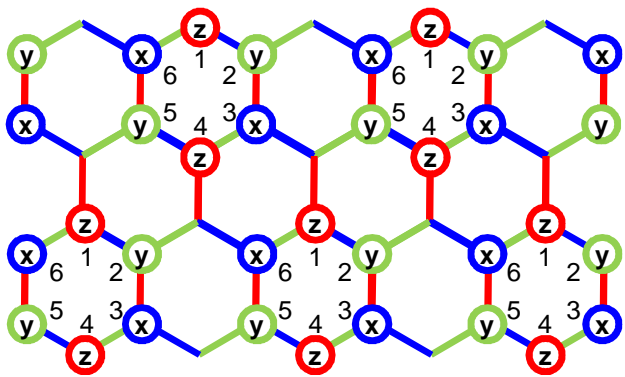}
\end{center}
\caption{(color online) The {\it efficient} pulse distribution 
$P_{\rm efc }$
for $H_{R}^{\rm efc}=P_{\rm efc }^\dagger H_SP_{\rm efc }$
in order to dynamically produce a Kitaev Hamiltonian 
from the Heisenberg model via one step. The $x$, $y$ and $z$ on the lattice sites 
show the application of $\pi/2$-pulses around $x$, $y$ and $z$, respectively.
} 
\label{efficient}
\end{figure}

{\it Efficient method.---}
The Kitaev Hamiltonian $H_K$ is produced more efficiently 
from $H_S$ when we apply rotation pulses more compactly.
Figure~\ref{efficient} shows the distributions of the rotation pulses
by which the BCH formula is used only once, such that $2\tau H_K^{\rm efc} =\tau (H_S + H_{R}^{\rm efc})$.
%We apply $\pi$-rotations around the $x$ and $y$ axes, depending on the qubit position 
%other than the qubits marked by the white circles.
The $x$-link of the colored honeycomb 
is produced by applying a rotation around the $y$-axis and that around the $z$-axis
on both sides of the link. Similarly, the $y$ ($z$)-link is produced by 
a rotation around the $z$ ($x$)-axis and around the $x$ ($y$)-axis on both sides of the link.
%Thus, if we consider an appropriate pulse arrangement, 
%we can produce the target Kitaev Hamiltonian by applying once the BCH formula. 

If $\tau_{\rm rot}$ denotes the time of a single-qubit rotation, 
it takes $2(2\tau_{\rm rot}+\tau)$ and $12(2\tau_{\rm rot}+\tau)$
to create rotations $\exp{ (-i2\tau H_K^{\rm efc})}$, 
and $\exp{ ( -i4\tau H_K^{\rm std})}$, respectively.
Similarly to the conventional DRAM, when we 
define {\it refresh overhead} as the effectiveness of the 
refresh of the quantum state such as
\begin{equation}
\text{refresh overhead} = \frac {\text{time required for refresh}} {\text{refresh interval}}.
\end{equation}
The refresh overhead of the efficient method presented above is 
$ (2\tau_{\rm rot}+\tau)/\tau \approx 2J_z\tau_{\rm rot}+1 $, 
and that of the standard method shown previously is $ 3(2J_z \tau_{\rm rot}+1)$, for $J_z \tau \lesssim 1$.
Thus, the efficient method is three times efficient than the standard method.
%%%%%%%%%%%%%%%%%%%%%%%%%%%%%%%%%%%%%%%%%%%%

{\it Fidelity.}---
%\section{Fidelity}
%The reduction of the number of the usage of  BCH formula 
%reduces the effects of the unwanted terms $H_{\rm uw}^{\rm bch}$ in the BCH formula 
%and improve the fidelity of the target Hamiltonian.
Let us numerically estimate the improvement of the efficient method
by calculating a {\it gate fidelity}~\cite{Daniel}. 
The time-dependent gate fidelity is defined by
\begin{equation}
F(t)=| {\rm Tr}[ \exp [it H_K] U_{\rm P}(t)] | /2^N,
\end{equation}%%%%%% check the definition!!
where 
$U_{\rm P}(t)$ denotes the evolution operator of the 
pulsed system. The gate fidelity shows how well the 
transformed Hamiltonian evolves compared with $H_K$.
For the standard arrangement, 
$U_{\rm P}(t)$ is given by 
$U_{\rm P}(t) 
=U_{\rm x}^{(1)}(t) U_{\rm x}^{(2)}(t)
 U_{\rm y}^{(1)}(t) U_{\rm y}^{(2)}(t)
 U_{\rm z}^{(1)}(t) U_{\rm z}^{(2)}(t)
$, 
with $U_{\rm \alpha}^{(i)}(t)=\exp [-it H_S]R_\alpha^{(i)} \exp [-it H_S]R_\alpha^{(i)}$, 
for $i=1,2$ and $\alpha=x,y,z$. 
$R_\alpha^{(1)}$ and $R_\alpha^{(2)}$ transform $H_S$ into 
the rotated ones shown in Figs.~\ref{standard}(a,c,e) and (b,d,f), respectively.
In contrast to this, for the efficient pulse arrangement, 
$U_{\rm P}(t)$ is simply expressed by $U_{\rm P}(t) 
=\exp [-it H_S]\exp[-it H_R]$. 

\begin{figure}
\begin{center}
\includegraphics[width=8.5cm]{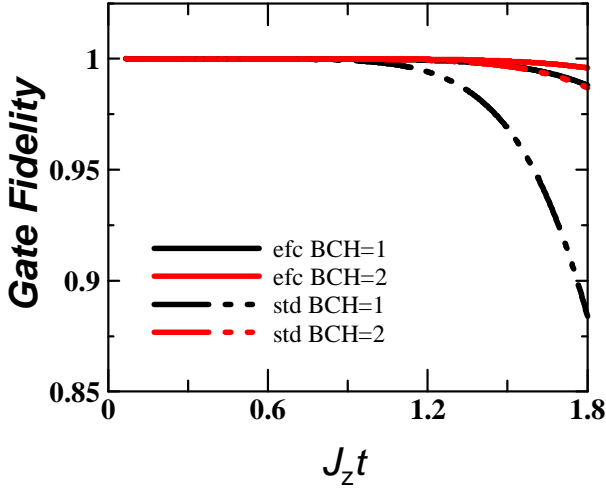}
\end{center}
\caption{(color online) Numerically calculated gate fidelity. 
Here ``std" corresponds to the $standard$ method (Fig.~\ref{standard}),
and ``efc" corresponds to the {\it efficient} method (Fig.~\ref{efficient}), respectively.
``BCH--$n$" means that the BCH formula is applied $n$ times.
Repeating the BCH formula corresponds to a
{\it refresh process}, which improves the gate fidelity. 
}
\label{Fidelity}
\end{figure}

Here the SO interaction and the HF interaction 
are included. % with constant coefficients $d_i$ and $h_i$ ($i=x,y,z$).
%In general spin qubits such as QD spin qubit, we cannot neglect 
%the SO interaction and the hyperfine interactions. 
The SO interaction is expressed by 
$V_{\rm so}=\sum_{jk} [ {\bf c}_{\rm so} \cdot 
(\mbox{\boldmath $\sigma$}_j-\mbox{\boldmath $\sigma$}_k) 
+{\bf d}_{\rm so} \cdot \mbox{\boldmath $\sigma$}_j \times \mbox{\boldmath $\sigma$}_k]$, 
where $\mbox{\boldmath $\sigma$}_j=(X_j,Y_j,Z_j)$, and the magnitudes of the spin-orbit vectors ${\bf c}_{\rm so}=(c_x, c_y, c_z)$ and 
${\bf d}_{\rm so}=(d_x, d_y, d_z)$ are 10$^{-2}$ smaller than $J_z$~\cite{Baruffa}.
The HF interaction is given by the fluctuation of the 
field such as $V_{\rm hp}=-\sum_j (\delta h_x X_j +\delta h_y Y_j +\delta h_z Z_j)$~\cite{Cywinsli}.
We treat the hyperfine field as a static quantity because the evolution of the hyperfine 
field is $\sim$10 $\mu$s and slower than the time scale of the pulse control 
$\sim$ 100~ns~\cite{Petta}.
The total Hamiltonian of this system in the calculation is expressed by
$ %\begin{equation}
H=H_S+V_{\rm so}+V_{\rm hp}.
$ %\end{equation}
The Chebyshev expansion method is used for calculating 
the time-dependent behavior until its 6th-order term~\cite{Kosloff}.
Figure~\ref{Fidelity} shows the numerical results for $N=10$ qubits (two honeycomb lattices). 
The calculations include several cases for different parameters 
(i) $J_x=J_y=0.3J_z$, $d_\alpha=0$, and $\delta h_\alpha=0$,   
(ii) $J_x=J_y=0.3J_z$, $d_\alpha=0.1$, and $\delta h_\alpha=0.1$, and  
(iii) $J_x=J_y=J_z$, $d_\alpha=0.3$, and $\delta h_\alpha=0.3$ 
($\alpha=x,y,z$).
%The effect of the difference of these parameters is very small, 
%and monitoring the gate fidelity, 
In various parameter regions, the overlap with 
the Kitaev Hamiltonian by using the pulse-controlled method is excellent.
We can also find that the repetition of the BCH formula~\cite{tanaSW} greatly 
increases the gate fidelity. In Fig.~\ref{Fidelity}, 
`BCH--2' means that two identical BCH operations
are repeated in a given time $t$ 
(See also Sec.I of \cite{supplement}).  
%%%%%%%%%%%%%%%%%%%%%%%%%%%%%%%%%%
% How do you treat ????
%%This repetition corresponds to the 
%%refresh process of the quantum state.
%%%%%%%%%%%%%%%%%%%%%%%%%%%%%%%%%%%
The repetition of the same operation also leads to the bang-bang control 
and reduces the effect of the noisy environment~\cite{Viola}.
The same feature is confirmed for a single honeycomb lattice $(N=6)$.
Thus we can say that the dynamical method on the Heisenberg 
model can realize the Kitaev Hamiltonian. 
%%%%%%%%%%%%%%%%%%%%%%%%%55555

\begin{figure}
\begin{center}
\includegraphics[width=8.5cm]{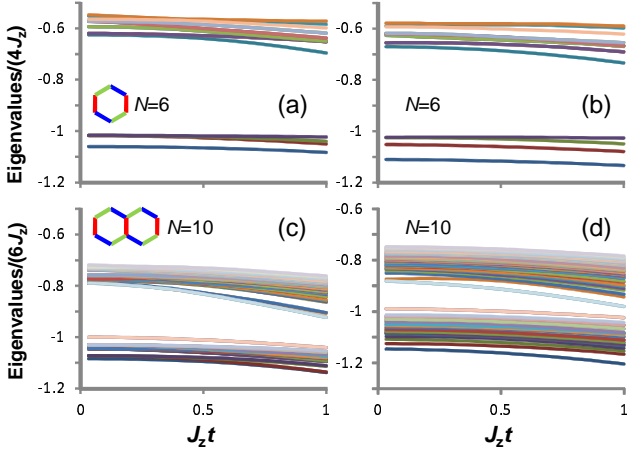}
\end{center}
\caption{(color online) Time-dependent eigenvalues of the effective Hamiltonian 
$H_{\rm eff}=\{ H_S +H_R -i(t/2)[H_S,H_R]\}/2$ for 
$N=6$ (a,b) and $N=10$ (c,d). $J_x=J_y=0.3J_z$
(a,c) use $d_x=d_y=\delta h_x=\delta h_y=0$.
(b,d) use $d_x=d_y=\delta h_x=\delta h_y=0.1J_z$.
%%%%%%%%%%% check !!
Eigenenergies are scaled by $J_z$.} 
\label{Eigenvalues}
\end{figure}
In order to directly see the effects of the unwanted terms, 
SO terms, and HF terms,
we calculate the time-dependent eigenvalues of the effective Hamiltonian 
$H_{\rm eff}=\{ H_S +H_R -i(t/2)[H_S,H_R]\}/2$, 
%$H_{\rm eff}=\{ H_S +H_R -i(t_0/4n)[H_S,H_R]\}/2$, 
as shown in Fig.~\ref{efficient}.
Because of limited computational resources, we 
show the numerical results for $N=6$ and $N=10$.
We can find that an energy gap opens up in the $J_z t \lesssim 1$ region.
The energy gap becomes narrow for $N=10$,
compared for $N=6$, because of finite-size effects.
When we compare Fig.~\ref{Eigenvalues}(d) with Fig.~\ref{Eigenvalues}(c), 
we find that the SO terms and the HF terms 
decrease the energy gap for large-$N$ systems.

%%%%%%%%%%%%%%%%%%%%%%%%%%%%%%%%%%%%%%%
{\it Toric code Hamiltonian.}---
%\section{Toric code Hamiltonian}
The unperturbed Hamiltonian of the $A$ phase is given by 
$H_0=-J_z \sum_{z-{\rm links}}Z_jZ_k$, whose ground state is 
a degenerate dimer state.
The Hamiltonian in the dimer state can be expressed by ``effective spin 
operators", $X^e$, $Y^e$ and $Z^e$, 
by pairing  original spin operators, such as
$P[X \times Y]\rightarrow  Y^e$,
$P[X \times X]\rightarrow  X^e$,
$P[Y \times Y]\rightarrow -X^e$,
$P[Z \times I]\rightarrow Z^e$, and
$P[Z \times Z]\rightarrow I^e$~\cite{Kells}.  
These pairs are taken between sites 2-3 and 5-6 in Fig.~1. The spins of the 
sites 1 and 4 are paired with the spins of another honeycomb lattice. 
Then $V_{\rm 0}=-J_x \sum_{x-{\rm links}}X_jX_k-J_y \sum_{y-{\rm links}}Y_jY_k$ 
acts as a perturbation and generates an effective Hamiltonian 
$H_{\rm eff}^{K}=-J_{\rm eff}^K \sum_p Q_p$, with 
$J_{\rm eff}^K =(J_x^2J_y^2/16J_z^3)$ and 
$Q_p=Y_{p,4}^eY_{p,2}^eZ_{p,1}^eZ_{p,4}^e$ 
in their forth-order effects. 
Therefore, here we have to compare $H_{\rm eff}^{K}$ 
with the perturbation terms in the BCH formula, the SO interaction, and
the HF interaction in the same framework as the Kitaev perturbation theory. 
It is obvious that the smaller magnitude 
of the SO terms and the HF interactions are desirable to achieve the 
condition that 
$ 
J_z> \{ J_x, J_y, |\vec{h}| \}> \{ |\delta \vec{h}_{\rm hf}|, c_{\rm so}, d_{\rm so}\}
$. Here we show that we have more constraints to realize the 
TQC. We also consider the commutation relation 
with a {\it plaquette operator} 
given by 
%\begin{equation}
$W_p=Z_1 Y_2 X_3Z_4Y_5X_6$, 
%\end{equation}
where $W_p$ commutes with the Hamiltonian Eq.~(\ref{Kitaev_H}), 
and is described by $W_p^e=Z^e_1Y^e_2Z^e_3Y^e_4$.
%'$A_z$ phase' where $J_z \gg J_x,J_y$. 

Let us start with the first-order unwanted terms in the BCH formula, 
given by 
$H_{\rm uw}=- it[H_S,H_R]/4$. 
%$H_{a1}=- it_0[H_S,H_R]/(4n)$. 
Many terms appear from this commutations relations (see Sec.IV of Ref.~\cite{supplement}). 
In order to see their typical effect, here we choose the unwanted terms 
that originate from the single honeycomb whose six qubits are rotated
(the center-bottom honeycomb in Fig.~\ref{efficient}).
These terms are explicitly expressed  such as
\begin{eqnarray}
\!\!\!&\! & \!H_{\rm uw}^{(l;z,x)}
=  (t_0/n)J_zJ_x[Y_{l}X_{l+1}Z_{l+2} - X_{l}Z_{l+1}Y_{l+2}],  \label{uw1}\\
\!\!\!&\! & \!H_{\rm uw}^{(l+1;y,z)}=  (t_0/n)J_zJ_y[X_{l+1}Z_{l+2}Y_{l+3}  -Z_{l+1}Y_{l+2}X_{l+3}], 
\ \ \ \ \label{uw2}\\
\!\!\!&\! & \! H_{\rm uw}^{(l+2;x,y)}=  (t_0/n)J_xJ_y[Z_{l+2}Y_{l+3}X_{l+4}  -Y_{l+2}X_{l+3}Z_{l+4}], 
\ \ \ \ \label{uw3}
%& & H_{a1}^{(123;z,x)}=  (t_0/n)J_zJ_x[Y_{1}X_{2}Z_{3} - X_{1}Z_{2}Y_{3}], \nonumber \\
%& & H_{a1}^{(234;y,z)}=  (t_0/n)J_zJ_y[X_{2}Z_{3}Y_{4}  -Z_{2}Y_{3}X_{4}], \nonumber \\
%& & H_{a1}^{(345;x,y)}=  (t_0/n)J_xJ_y[Z_{3}Y_{4}X_{5}  -Y_{3}X_{4}Z_{5}], \nonumber \\
%& & H_{a1}^{(456;z,x)}=  (t_0/n)J_zJ_x[Y_{4}X_{5}Z_{6} - X_{4}Z_{5}Y_{6}], \nonumber \\
%& & H_{a1}^{(561;y,z)}=  (t_0/n)J_zJ_y[X_{5}Z_{6}Y_{1}  -Z_{5}Y_{6}X_{1}], \nonumber \\
%& & H_{a1}^{(612;x,y)}=  (t_0/n)J_xJ_y[Z_{6}Y_{1}X_{2}  -Y_{6}X_{1}Z_{2}]. \label{HaHb}
\end{eqnarray}
for $l=1,4$. 
Because these terms do not commute 
with $W_p$ nor the Hamiltonian Eq.(\ref{Kitaev_H}),
 they are unwanted terms in the topological quantum computation.
The effective Hamiltonian of Eqs.(\ref{uw1})-(\ref{uw3}) 
appears in the second-order perturbation 
given by $\langle a | H_{\rm eff}^{(2)} |b\rangle 
=\sum_j' [\langle a | V |j\rangle \langle j | V |b\rangle/(E_0-E_j)]$, 
which is expressed as
\begin{eqnarray}
\!\! &\!  H_{\rm eff}^{\rm uw} \! &\!\! \approx  t^2 %\frac{t_0^2}{n^2}
\large\{
J_z
\left[J_x^2 (X_1^eX_2^e+X_3^eX_4^e)
%&+&
\!+\!J_y^2 (X_2^eX_3^e+X_4^eX_1^e) \right]
\nonumber \\
&+&J_x^2J_y (2Z_2^eZ_4^e-Z_1^eX_2^eZ_4^e-Z_3^eZ_2^eX_4^e)
\nonumber \\
&+&J_xJ_y^2 (2Z_2^eZ_4^e+Z_1^eZ_2^eX_4^e+Z_3^eX_2^eZ_4^e)
\large\}.
%+O(J..)
\end{eqnarray}
%\begin{eqnarray}
%\lefteqn{ H_{\rm eff}^{bch}\approx\frac{t_0^2}{n^2}
%\biggr\{
%\frac{(J_zJ_x)^2}{J_z} (X_1^eX_2^e+X_3^eX_4^e)}
%\nonumber \\
%&+&\frac{(J_zJ_y)^2}{J_z} (X_2^eX_3^e+X_4^eX_1^e)
%\nonumber \\
%&+&\frac{J_zJ_x^2J_y}{J_z} (2Z_2^eZ_4^e-Z_1^eX_2^eZ_4^e-Z_3^eZ_2^eX_4^e)
%\nonumber \\
%&+&\frac{J_zJ_xJ_y^2}{J_z} (2Z_2^eZ_4^e+Z_1^eZ_2^eX_4^e+Z_3^eX_2^eZ_4^e)
%\biggl\}.
%+O(J..)
%\end{eqnarray}
%%(Check lowest order commutation relation)
Thus, this term commutes neither with the Kitaev Hamiltonian nor with $W_p$. 
%%(Check exactly!).
From this approximation, to realize a TQC, 
we have a constraint on the time, given by
$t^2 J_\alpha ^2 J_z < J_{\rm eff}^K$. 
%%$(t_0/n)^2 J_\alpha ^2 J_z < J_{\rm eff}^K$. 
%  (t_0/n)^2 J_x ^2 J_z < J_{\rm eff}^K=Jx^2Jy^2/(16Jz^3)
%  (t_0/n)^2  < Jy^2/(16Jz^4)
%  (t_0/n)  < Jy/(4Jz^2)
%  t_0  < n Jy/(4Jz^2)
When $J_x=J_y$, this corresponds to 
$t <  J_x/(4J_z^2)$.
%%$t_0 < n J_x/(4J_z^2)$.
The effective SO terms are derived in the similar manner. 
There are many terms regarding the SO interactions (See Sec.VI of Ref.~\cite{supplement}),
all of which do not commute 
with $W_p$ nor the Hamiltonian Eq.(\ref{Kitaev_H}).
As an example, the effective SO terms of the center honeycomb lattice 
at the bottom row in Fig.~3 are given by
$H_{\rm eff}^{\rm so}=[2d_xd_y (X_2^e+X_4^e)/J_z]$. 
%\end{equation}
%% 
The effect of the HF interaction is expressed by 
$H_{\rm eff}^{\rm hp}=[\langle \delta h_{x2}\delta h_{y3}\rangle Y_2^e
+\langle \delta h_{x5}\delta h_{y6}\rangle Y_4^e]/J_z$, 
assuming the uniformity of the HF interaction, 
such as $\langle \delta h_{x2}\delta h_{x3}\rangle =\langle \delta h_{y2}\delta h_{y3}\rangle$. 
This term also does not commute 
with $W_p$ nor the Hamiltonian Eq.(\ref{Kitaev_H}). \color{black}
%\end{eqnarray}
From these estimates, in order to realize the TQC,
both the SO and the HF interactions should be small and 
we have the constraint 
$2d_xd_y/J_z$ and $\langle \delta h_{x5}\delta h_{y6}\rangle/J_z < J_{\rm eff}^K$.

{\it Discussion.}---
Let us consider a process of TQC in spin qubits. 
The toric code and the surface code are based on stabilizer formalism~\cite{Bravyi,Fowler},
where desired quantum states are obtained by stabilizer measurements.
These measurements can be carried out by conventional spin-qubit  operations 
by manipulating the Heisenberg model with appropriate magnetic fields.
%These measurements can be carried out by conventional spin-qubit  operations, and the %desired states can be obtained 
%in the range of the Heisenberg model.
%Even if we obtain the desired states by these measurements, 
%unless the desired states are not 
%the eigenstates of the system Hamiltonian, 
However, because the desired states are not always  
eigenstates of the Heisenberg Hamiltonian,
the desired states are not preserved. 
Thus, the present method which can preserve the desired states of the TQC
is important after the measurement. 
Let us estimate the measurement time 
from this scheme. \color{black}
In each measurement process of the surface code, 
four CNOT gates and two Hadamard gates are required~\cite{Fowler}. 
When each CNOT gate consists of two $\sqrt{\rm SWAP}$s~\cite{Burkard}
and each $\sqrt{\rm SWAP}$ requires a time $\pi/(8J_{\rm meas})$, 
where $J_{\rm meas}$ is a Heisenberg coupling strength for the measurement, 
one stabilizer measurement cycle approximately requires a time $\pi/J_{\rm meas}$.
Because a short measurement time and a long coherence-preserving time $(\sim J_z^{-1}$)
are preferable, it is desirable for
the coupling strength between qubits to be changeable, 
therefore $J_{\rm meas} > J_z$ is desirable. 

In  spin-qubit systems based on QDs, 
the coupling $J_{jk}$ exponentially changes 
as the distance between two QDs or 
the gate voltage changes~\cite{Petta,Xuedong,Burkard2}.
%by controlling the tunneling height 
%between the QDs by the gate voltages where 
As an example,  $J\approx 0.1$--$1~\mu$eV is obtained,  
when the voltage difference between two GaAs QDs 
is less than 10 mV~\cite{Petta}, and
we can choose $J_z\approx0.1~\mu$eV and $J_{\rm meas}\approx 1~\mu$eV.
When $J_z\approx 0.1~\mu$eV (= 0.0116 K), the period $J_zt \lesssim 1$ 
corresponds to $t \sim 24.2$ ns. 
In the typical DRAM array, every row is 
refreshed about every 15 $\mu$s~\cite{DRAM}. 
Thus, compared with the conventional DRAM, 
a more frequent refresh is required in our proposal.
%In order to use wide range of the coupling constants, 
%the SO and 
%HF interactions are expected to be as small as possible.
%When we consider that short pulses are not easy to  
%generate in many spin-qubit systems, then a smaller $J_z$ is 
%preferable. From this viewpoint, 

Experimentally, spin-qubit systems are realized by 
either Si donors or NV centers, in addition to semiconductor QDs, such as GaA or SiGe.
The SO and the HF interactions are 
strongest in GaAs systems and smallest 
in donor systems and NV centers. The controllability of spins is 
better for GaAs systems and difficult in donor systems and NV centers.
Thus there is a tradeoff between the controllability and 
the realization of TQC.

In summary, we proposed how to dynamically generate a Kitaev spin Hamiltonian 
on a honeycomb lattice from the Heisenberg spin Hamiltonian
by using a dynamical average Hamiltonian theory. 
We also considered the effects of the 
unwanted terms of the BCH, 
SO interaction, and HF interactions.
We clarified that if these terms are sufficiently small, 
a dynamic TQC is available by periodically 
reproducing the topological Hamiltonian.

\acknowledgements
TT would like to thank 
A. Nishiyama, K. Muraoka, S. Fujita, and H. Goto for discussions.
 YXL is supported by the National Natural Science Foundation of China under Grant Nos. 61025022, 91321208, the National Basic Research Program of China Grant No. 2014CB921401.
FN is partially supported by the 
RIKEN iTHES Project, 
MURI Center for Dynamic Magneto-Optics, 
and a Grant-in-Aid for Scientific Research (S). 

%############################################333
%%%%%%%%%%%%%%%%%%%%%%%%%%%%%%%%%%%%%%%%%%%%%%

\end{document}